\documentclass[pra,twocolumn,superscriptaddress,nobibnotes,letterpaper]{revtex4-1}

\usepackage[english]{babel}
\usepackage[utf8]{inputenc}
\usepackage{chemformula,amsmath,amsfonts,graphicx,siunitx}
\usepackage[colorinlistoftodos]{todonotes}
\usepackage[colorlinks=true, allcolors=blue]{hyperref}

\newcommand{\at}{\makeatletter @\makeatother}

\begin{document}

\title{Integrated optomechanical arrays of two high reflectivity SiN membranes}\thanks{This work was published in Nano Lett.\ \textbf{18}, 7171 (2018).}

\author{Claus G\"artner}\thanks{These authors contributed equally to this work.}
\affiliation{Vienna Center for Quantum Science and Technology (VCQ), Faculty of Physics, University of Vienna, A-1090 Vienna, Austria}
\affiliation{Department of Quantum Nanoscience, Kavli Institute of Nanoscience, Delft University of Technology, Lorentzweg 1, 2628CJ Delft, The Netherlands}
\author{Jo\~{a}o P.\ Moura}\thanks{These authors contributed equally to this work.}
\author{Wouter Haaxman}
\author{Richard A. Norte}
\author{Simon Gr\"oblacher}
\email{s.groeblacher@tudelft.nl}
\affiliation{Department of Quantum Nanoscience, Kavli Institute of Nanoscience, Delft University of Technology, Lorentzweg 1, 2628CJ Delft, The Netherlands}

\begin{abstract}
Multi-element cavity optomechanics constitutes a direction to observe novel effects with mechanical resonators. Several exciting ideas include superradiance, increased optomechanical coupling, and quantum effects between distinct mechanical modes among others. Realizing these experiments has so far been difficult, because of the need for extremely precise positioning of the elements relative to one another due to the high-reflectivity required for each element. Here we overcome this challenge and present the fabrication of monolithic arrays of two highly reflective mechanical resonators in a single chip. We characterize the optical spectra and losses of these \SI{200}{\micro m}-long Fabry-P\'{e}rot interferometers, measuring finesse values of up to \num{220}. In addition, we observe an enhancement of the coupling rate between the cavity field and the mechanical center-of-mass mode compared to the single membrane case. Further enhancements in coupling with these devices are predicted, potentially reaching the single-photon strong coupling regime, giving these integrated structures an exciting prospect for future multi-mode quantum experiments.
\end{abstract}

\maketitle

Cavity optomechanics explores light-matter interactions by using the established control techniques of optical resonators to manipulate highly sensitive mechanical oscillators~\cite{Aspelmeyer2014}. A particularly successful direction is to dispersively couple suspended silicon nitride (SiN) membranes to a rigid optical cavity~\cite{Thompson2008}. These so called membrane-in-the-middle (MIM) systems combine independent optical and mechanical oscillators, allowing the use of high finesse cavities to study a variety of mechanical devices. Although recent years have seen tremendous progress in quantum optomechanics and in particular with experiments observing quantum behavior of the mechanical mode~\cite{OConnell2010,Hong2017,Riedinger2018}, most have focused on single mechanical or noninteracting modes. Studying the behavior of multiple directly coupled modes could however allow probing new and exciting regimes of optomechanics~\cite{Bhattacharya2008}, like superradiance, phonon lasing~\cite{Gross1976,Kipf2014}, synchronization~\cite{Zhang2012}, the study of exceptional points~\cite{Xu2016}, quantum information processing~\cite{Schmidt2012}, as well as the direct entanglement of mechanical resonators~\cite{Hartmann2008}. It has also been suggested that the collective interaction of several mechanical oscillators can allow the reaching of the single-photon strong coupling regime~\cite{Xuereb2012}. This effect is based on reducing the effective optical mode volume through an array of closely spaced mechanical systems and it becomes stronger as the reflectivity of the individual systems $R_m$ is increased.

Tethered SiN membranes patterned with photonic crystals (PhC) constitute ideal candidates for this type of experiments, as they have excellent mechanical properties, low mass, and high reflectivity due to the PhC which can be engineered to operate at a large range of wavelengths~\cite{Norte2016,Reinhardt2016}. To date, experimental efforts have focused on using independent mechanical membranes to create a mechanical array~\cite{Nair2017,Piergentili2018}, relying on the intrinsic reflectivity of the bare SiN with one recent attempt to fabricate a membrane on each side of the same chip~\cite{Weaver2017}.

In the present work, we monolithically combine two tethered SiN membranes on a single chip and control their reflectivity using PhC patterns. This allows us to avoid having to manually align the mechanical elements to each other, which to date has been a major challenge with such high-reflectivity resonators. To compare the properties of devices with different reflectivity $R_m$, we fabricate pairs of single and double-membranes for three different PhC parameter sets, spanning $R_m$ from \SI{33}{\percent} to \SI{99.8}{\percent} at an operating wavelength of \SI{1550}{nm}. The optical spectrum of the arrays exhibits Fabry-P\'{e}rot interference, which allows us to study the optical loss mechanisms present in the system. The optomechanical coupling rate of the center-of-mass (COM) mode of single and double-membranes to an optical cavity are compared. By changing the incident laser wavelength, we can operate the double-membrane stacks in their reflective or transmissive regimes, corresponding to enhanced or null COM optomechanical couplings, respectively.

\section*{Device design and fabrication}

\begin{figure}[t]
  \centering
  \includegraphics[width=1\columnwidth]{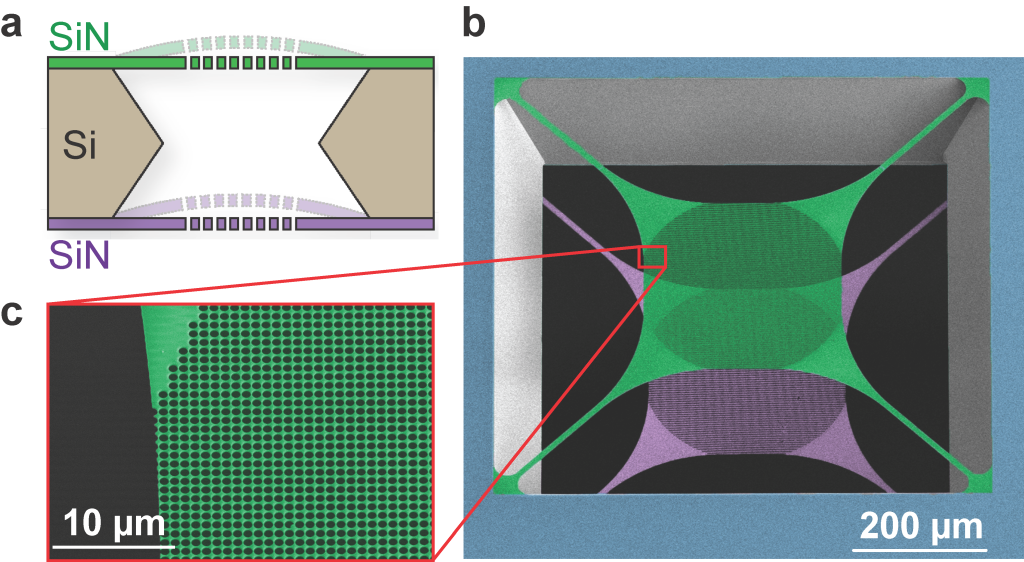}
  \caption{\textbf{a} Cross-sectional schematic of a released double membrane stack. \textbf{b} False-colored SEM image from the top under an angle of \ang{37} showing a stack of two membranes as depicted in \textbf{a}. The top (green) and bottom (purple) SiN trampolines form a Fabry-P\'{e}rot cavity. \textbf{c} Zoom-in of the PhC patterned central pad area of the upper membrane.}
  \label{fig:device}
\end{figure}

We fabricate our optomechanical devices on \SI{200}{nm} of low-pressure chemical-vapor deposition (LPCVD) SiN deposited on both sides of a \SI{200}{\micro m} thick silicon (Si) substrate. A trampoline membrane is patterned on each side of the chip using electron-beam lithography and then etched into the SiN using a \ch{CHF3}/\ch{O2} plasma etch. Finally, the Si in-between the trampolines is removed with KOH etching. Figure~\ref{fig:device} shows a cross-sectional schematic of a final double membrane stack, as well as a false-colored SEM of one of our released devices.

At the heart of our devices is a central mirror pad on the tethered membranes. It is patterned with a two-dimensional PhC consisting of a periodic array of holes etched into the SiN device layer. Such a periodic change in the refractive index creates a band gap that can be tailored to a specific wavelength, resulting in reflectivities $> \SI{99.9}{\percent}$~\cite{Norte2016,Chen2017}. Using $S^4$, a Rigorous Coupled-Wave Analysis software, we simulate the spectrum of a given PhC pattern~\cite{LiuFan2012}. During fabrication, we can accurately tune the PhC resonance to our desired wavelength by adjusting the lattice constant $a$ and hole radius $r$ (see Figure~\ref{fig:Tuning} in the Supplementary Information for more details). We design three PhC patterns in order to obtain different $R_m$ at our operating wavelength of \SI{1550}{nm}. We refer to these patterns as \emph{Low}, \emph{Mid}, and \emph{High $R$} and their geometries and measured $R_m$ at \SI{1550}{nm} are specified in Table~\ref{table:PhC}. The optical beam we use to probe the PhC has a waist size of about \SI{50}{\mu m}. To avoid clipping losses, the diameter of the PhC pattern is \SI{300}{\mu m}, while the tether length and width are \SI{318}{\mu m} and \SI{10}{\mu m}, respectively.

\begin{figure*}[t]
  \centering
  \includegraphics{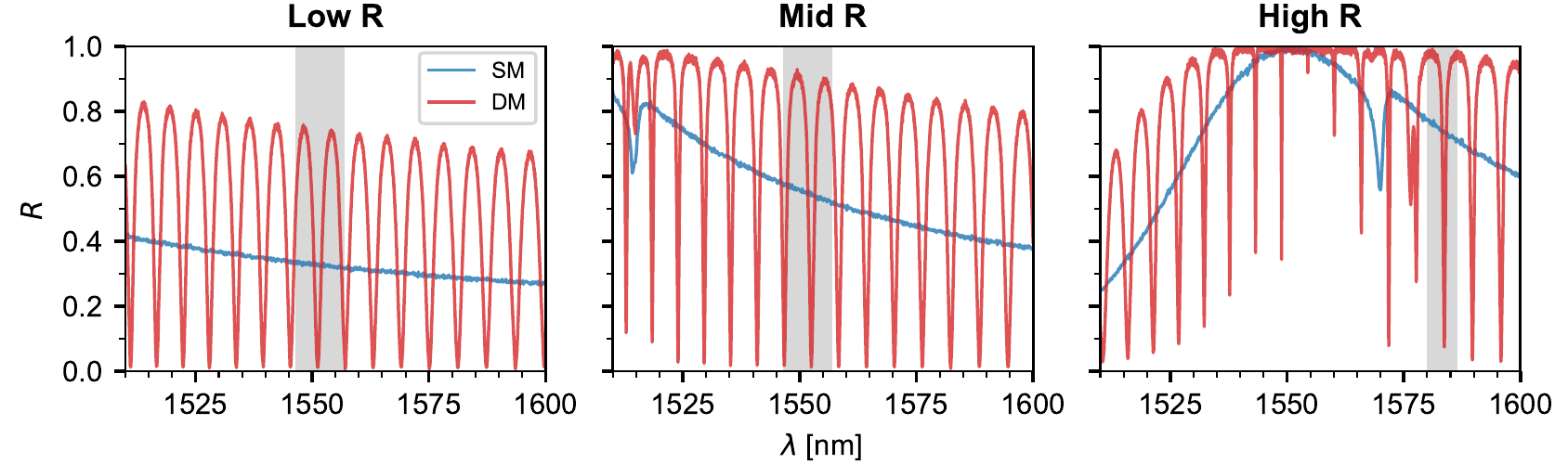}
  \caption{Reflection spectra of the devices. The photonic crystal pattern of each device is indicated at the top of each figure, according to the parameters in Table~\ref{table:PhC}. The blue traces correspond to devices composed of a single-membrane (SM), whereas the red traces are from double-membrane stacks (DM). The gray-shaded regions correspond to the wavelength ranges where the center-of-mass optomechanical coupling was measured (see Figures~\ref{fig:dispersion} and \ref{fig:coupling}).}
  \label{fig:transmission}
\end{figure*}

\section*{Results and discussion}

For each PhC pattern (cf.\ Table~\ref{table:PhC}), we fabricate a single- and a double-membrane, which allows us to test all designs on a single chip, greatly facilitating the measurements. In the following subsections, we characterize their optical, mechanical, and optomechanical properties.

\subsection*{Optical characterization}

\subsubsection*{Single membranes}
We first obtain the optical spectra of the single devices by scanning a tunable laser from \num{1510} to \SI{1600}{nm} and measure the reflected and transmitted signals from the PhC trampolines, which are shown in Figure~\ref{fig:transmission}. At \SI{1550}{nm}, we measure reflectivities of \SI{33}{\percent}, \SI{56}{\percent}, and \SI{99.8}{\percent} for the Low, Mid, and High R samples, respectively. Because this measurement procedure has an uncertainty of \SI{0.5}{\percent}, we determine the dispersive effect of a device similar to the High R sample on an optical cavity to obtain a lower bound on its transmission at resonance~\cite{Stambaugh2014,Chen2017}. We measure a transmission of \num{2.5e-5}, comparable to the best reported results in the literature~\cite{Chen2017}. Finally, we simulate a PhC membrane with an imaginary component of the refractive index of \num{1.9e-5}~\cite{Stambaugh2014}, and estimate that a fraction of \num{3.4e-4} of the light is lost when interacting with the devices, due to either absorption or scattering from fabrication imperfections (see Supplementary Information for more details).

\begin{table}[t!]
	\begin{tabular}{c|ccc}
		& $a$ [nm] & $r$ [nm] & $R_m$\at\SI{1550}{nm} \\ \hline
		Low R          & 1240            & 475             & \SI{33}{\percent}\\
		Mid R          & 1310            & 500             & \SI{56}{\percent}\\
		High R         & 1372            & 525             & \SI{99.8}{\percent}
	\end{tabular}
	\caption{Lattice constant $a$ and hole radius $r$ of the PhC patterns used in this work, as well as their measured reflectivity $R_m$ at our operating wavelength of \SI{1550}{nm}.}
	\label{table:PhC}
\end{table}

\subsubsection*{Double-membrane arrays}

The double-membrane arrays have the same PhC design as the individual membranes and we determine their optical response in a similar way, shown in Figure~\ref{fig:transmission}. These structures can be modeled as plane-parallel etalons (Fig.~\ref{fig:device}) and the characteristic features of Fabry-P\'{e}rot interferometers can be clearly observed in their spectra. The free spectral range $\rm FSR_{DM}$ of \SI{750}{GHz}, or \SI{6}{nm} at a wavelength of \SI{1550}{nm}, is, as expected, defined by the \SI{200}{\mu m} thickness of the Si substrate that separates the two membranes. The linewidth of the resonances becomes smaller as the reflectivity of the individual membranes increases. This is particularly prominent on the High R sample, where the full-width at half-maximum linewidth changes from \SI{176}{GHz} at \SI{1521}{nm} to \SI{8.7}{GHz} at \SI{1554}{nm}, corresponding to a change in finesse $F$ from \num{4.3} to \num{86}. Our best performing samples exhibit linewidths as low as \SI{3.3}{GHz} ($F=220$), suggesting a total loss per round-trip of approximately $2\pi/F = \num{2.9e-2}$.

Several sources contribute to this loss. First, using the measurements presented in the previous section, we estimate a lower bound for the round-trip transmission of \num{5e-5}. However, in general the highest finesse etalon peak is not exactly at the resonance of the PhC, being at most $\rm FSR_{DM}/2=\SI{3}{nm}$ away from it. At this point, the round-trip transmission becomes \num{2.6e-2}. Second, we expect a round-trip absorption and scattering loss of \num{6.8e-4}. Finally, some light will be lost due to the finite aperture size of the etalon. Plane-parallel Fabry-P\'{e}rot cavities are particularly susceptible to this effect~\cite{Siegman1986,Svelto2010}, and we estimate it to result in a round-trip loss of \num{2e-3}. Combining these effects we arrive at estimated total round-trip losses between \num{2.8e-3} and \num{2.9e-2} (see Supplementary Information for more details).

Although the maximum finesse measured in our devices fits well to this range, the fact that we generally measure lower values suggests that they are underestimated. Scattering, which has consistently been identified as one of the main loss mechanisms in other PhC membranes \cite{Stambaugh2014,Chen2017}, could be higher than expected. In addition, these estimates assume that both membranes have the same reflectivity. In both the Low and Mid R samples the reflection drops to zero at the etalon resonances, indicating that the PhC resonances on the front and back membranes are sufficiently well matched in these regimes. However, with increasing reflectivities, mismatches due to fabrication imperfections and small systematic shifts between the individual PhC mirrors become more apparent and lead to smaller dip depths (cf.\ the High R device in Figure~\ref{fig:transmission}). In fact, as the reflectivity of the individual membranes increases, the dip depth becomes significantly more sensitive to differences between the two mirrors (see Figure~\ref{fig:Impedance} in the Supplementary Information). This also results in higher round-trip transmission values that can explain the discrepancy between our finesse estimates and measurements.

\subsection*{Mechanical characterization}

We determine the mechanical quality factor of the fundamental modes of both single and double membrane devices by performing interferometric ring-down measurements. The mode frequencies are approximately \SI{150}{kHz} and the difference in frequency between the front and back membranes is typically around \SI{170}{Hz}. The small difference of around \SI{0.1}{\percent} in resonance frequency can be attributed to an irreproducibility in the fabrication process. All devices show unclamped quality factors between \num{1.2e6} and \num{5.6e6}. These values are in good agreement with measurements on a similar geometry, which showed quality factors of \num{4e6}~\cite{Norte2016}, indicating that the PhC patterning does not negatively effect their mechanical properties.

\begin{figure*}[t]
  \includegraphics[width=1.9\columnwidth]{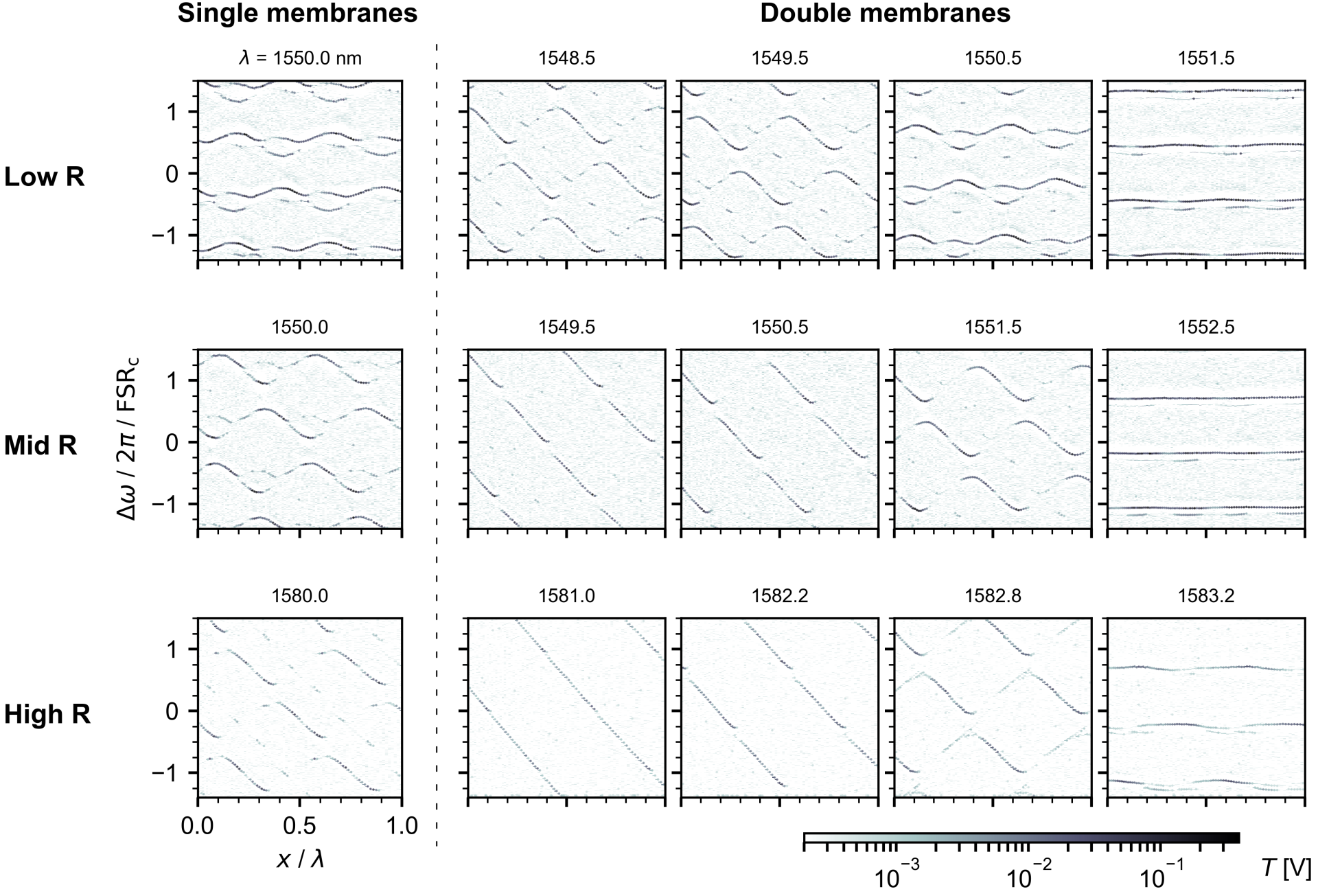}
  \caption{Optical cavity transmission $T$ as a function of the frequency shift $\Delta\omega$ of the incident laser and of the displacement $x$ of several mechanical devices placed in the middle of the cavity. $\Delta\omega$ is normalized by the cavity free spectral range ${\rm FSR_c}=\SI{3.13}{GHz}$ and $x$ by the laser wavelength $\lambda$ which is indicated on top of each plot. We measured multiple devices in the middle of the cavity:\ on the left of the dashed line we study single-membranes and on the right double-membranes. The type of photonic crystal used in each sample is indicated on the left of the figure. Note that in order to work in a regime with a slow reflectivity change and large dip depth, the High R samples were studied at a wavelength for which $R_m=\num{0.76}$.}
  \label{fig:dispersion}
\end{figure*}

\subsection*{Optomechanical characterization}

In order to obtain the optomechanical characteristics of the devices we place them inside an optical cavity. The optical modes of this larger cavity strongly depend on the position of the membranes inside. By measuring the changes in cavity mode frequency $\omega_c$ as a function of the device displacement $x$, we are able to determine the linear optomechanical coupling between the cavity and the device's center-of-mass mechanical modes, which we define as $G\equiv\max\{\left|\partial \omega_c/\partial x\right|\}$. The cavity has a free spectral range ${\rm FSR}_c = \SI{3.13}{GHz}$ and an empty cavity half-width at half-maximum of $\kappa/2\pi = \SI{550}{kHz}$. We align our tunable laser to the cavity and measure the transmitted light. The laser frequency is then scanned as a function of the device position, which allows us to directly obtain $\omega_c(x)$ and calculate the optomechanical coupling.

Let us first consider the case of a single-membrane, where the cavity modes are affected by the membrane position and reflectivity $R_m$, according to $\Delta\omega_c/2\pi ={\rm FSR}\cdot\arccos(\sqrt{R_m}\cos(4\pi x/\lambda))/\pi$~\cite{Thompson2008}. The so-called linear coupling regime occurs when a membrane is placed close to $x = \lambda/8 + n\lambda/4,n\in \mathbb{Z}$. Around these points, the cavity frequency changes linearly with the membrane position and the optomechanical coupling is given by
\begin{equation}
\label{eq:dispersion-SM} \frac{G}{2\pi} = 4 \frac{\rm FSR}{\lambda}\sqrt{R_m}.
\end{equation}

\begin{figure}[t]
  \includegraphics{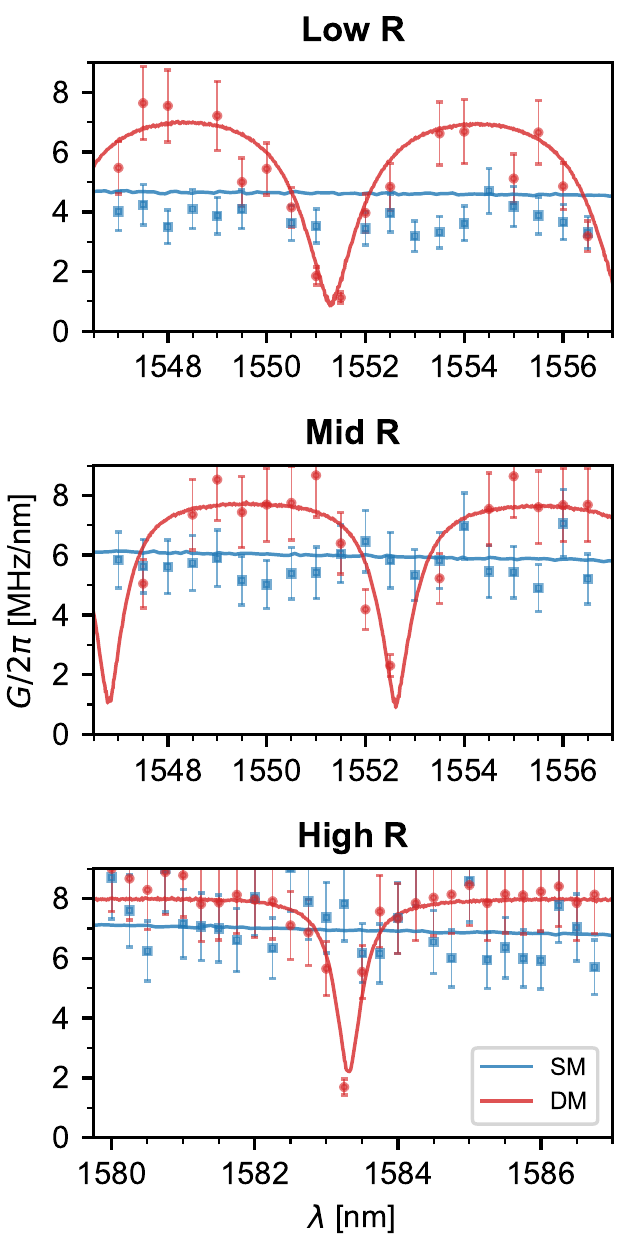}
  \caption{Center-of-mass optomechanical coupling $G/2\pi$ as a function of wavelength $\lambda$ obtained through the derivative of the cavity dispersion $\max\{\left|\partial \omega_c/\partial x\right|\}$ (points) and through the membrane-in-the-middle model $4 \frac{\rm FSR}{\lambda}\sqrt{R_m}$ (lines). The blue data are taken from single- (SM) and the red from double-membrane (DM) devices. The corresponding PhC patterns are indicated on top of each figure with the wavelength range studied here marked in gray in Fig.~\ref{fig:transmission}.}
  \label{fig:coupling}
\end{figure}

The first column of Figure~\ref{fig:dispersion} shows the cavity transmission as a function of laser frequency shift and displacement of the single-membrane samples. The wavelength at which the measurements were taken is indicated above each plot. The points of high transmission correspond to cavity modes. Because of alignment imperfections between the laser, the cavity and the membranes, in addition to the fundamental cavity mode, we also observe higher order modes, which can be coupled to each other~\cite{Sankey2010}. The fundamental optical mode frequency depends on the membrane position with a periodicity of $x/\lambda = \pi/2$ and the amplitude of the frequency oscillations increases with the membrane reflectivity, as indicated by Eq.~\eqref{eq:dispersion-SM}. Using these data, we obtain $G$ by numerically calculating $\left|\partial \omega_c/\partial x\right|$ and taking its maximum value, which occurs at the positions of linear coupling. The blue data points in Figure~\ref{fig:coupling} show the single membranes' coupling around a narrow wavelength window. In addition we plot the coupling as calculated by the reflectivity measured in Fig.~\ref{fig:transmission} and Eq.~\eqref{eq:dispersion-SM}. Within this wavelength range, the reflectivity of each device varies little and therefore $G$ is practically constant. The average measured couplings $G/2\pi$ for the Low, Mid, and High R samples are \num{3.8\pm0.6}, \num{5.7\pm0.9} and \num{7.7\pm1.2}~MHz/nm, whereas the expected values using Eq.~\eqref{eq:dispersion-SM} and $\sqrt{R}$ are \num{4.5}, \num{5.8}, and \num{6.8}~MHz/nm. Despite the large uncertainty, mainly due to the displacement calibration, the results are in good agreement with Eq.~\eqref{eq:dispersion-SM}.

Finally, we follow the same approach to obtain the coupling rate between the cavity and the COM displacement of the double-membrane chips, schematically represented in Fig.~\ref{fig:device}. The crucial difference between single and double-membranes is that the latter's spectra vary more strongly with wavelength. In particular, over one $\rm FSR_{DM}$, the device reflectivity can quickly change from zero to one (see Fig.~\ref{fig:transmission}). When the reflectivity is low, the COM mode of the device will interact weakly with the external cavity. Correspondingly, at a reflection maximum, the coupling will be higher than that of a device composed of only one membrane. In columns 2 to 5 of Fig.~\ref{fig:dispersion} the measured cavity dispersion for the three double-membranes studied is shown. We perform these measurements at several wavelengths spanning half a $\rm FSR_{DM}$, between which the reflectivity varies between its maximum and minimum values. Note that for the High R sample we choose to study a resonance for which $R_m \sim 0.76$ ($\lambda$ close to \SI{1580}{nm}) since for higher $R_m$ the laser fine scanning range becomes similar to the resonance linewidth, and the dip depth decreases, making the coupling oscillations less visible. Column 2 corresponds to the reflectivity maxima. When comparing it with column 1, it becomes clear that the cavity frequency varies more strongly than in the single-membrane case. The data in column 5 are taken close to a transmission maximum where, as discussed, the COM motion has little influence on the cavity frequency. Columns 3 and 4 show wavelengths in between the maximum reflection and transmission of the double-membrane stacks. The extracted COM coupling is plotted in red in Fig.~\ref{fig:coupling}. As discussed, the coupling oscillates between almost zero and values larger than those of the individual membranes. The oscillation follows the device's spectral response, indicating that the COM coupling of a double-membrane is well described by Eq.~\eqref{eq:dispersion-SM}, a model derived from the single-membrane case.

In conclusion, we have fabricated and characterized stacks of optomechanical devices that operate in various low to high reflectivity regimes. The devices presented here are patterned onto a single chip without the need for additional bonding steps or micro-positioners. Our devices form a flexible platform in which the finesse can be freely tuned. Placing these devices inside an optical cavity allows the direct comparison of membrane-in-the-middle systems in multiple reflectivity regimes, such as proposed in Ref.~\cite{Xuereb2012NJP}. We see an enhancement of the optomechanical coupling rate between the COM motion of the two membranes and the cavity field as a function of reflectivity, when compared to a single membrane system.

More importantly, we can tune the system such that the COM coupling is practically zero. The theory of the collective motion of optomechanical arrays predicts that at these points the cavity field becomes resonant with the inner cavity and thus couples strongly to the relative motion of the membranes. This is the regime where single-photon strong coupling in an optomechanical system could be achievable~\cite{Xuereb2012}. We are currently working on improving the stability of our setup in order to probe these relative motional modes. For devices with large $R_m$, like the ones presented here, the coupling enhancement of the differential mechanical motion is limited by the ratio $L/2d$ between the length of the optical cavity $L$ and the separation between the membranes $d$~\cite{Li2016}. Given our experimental parameters, this should allow us to observe an enhancement of up to 120. Increasing this value further could be done by replacing the Si substrate by a thin sacrificial layer as the spacer between mirrors, considerably decreasing $d$ to values similar to~\cite{Nair2017} but keeping the advantages of monolithic fabrication presented here.

Even more interestingly, the single-photon cooperativity scales quadratically with the single-photon coupling strength, which in our case could boost this important figure of merit by 4 orders of magnitude, assuming the mechanical and optical dissipation rates stay the same. For many experiments, coherent control in the strong single-photon coupling regime is not necessary but reaching cooperativities greater than one is sufficient for performing several quantum protocols~\cite{Aspelmeyer2014,Leijssen2017}. Other interesting experiments could include synchronization of mechanical modes~\cite{Zhang2012}, studying exceptional points in optomechanics with independent mechanical systems, as well as superradiance~\cite{Gross1976,Kipf2014} and state transfer between mechanical systems~\cite{Weaver2017}. In addition, our arrays could serve as rigid, stable free-space optical filters with adjustable finesse. The arrays also constitute an optomechanical system by themselves, whose mirrors are both movable and with engineerable optical and mechanical properties. As both mirrors and mechanical resonators are monolithically combined, the system is inherently stable, greatly relaxing the setup complexity of typical free-space optomechanical setups, and making it an ideal platform for simple studies of radiation-pressure effects.

\section*{Acknowledgments}

We would like to thank Markus Aspelmeyer and Clemens Sch\"afermeier for discussions and support. We also acknowledge assistance from the Kavli Nanolab Delft, in particular from Marc Zuiddam, Charles de Boer, and Arnold van Run. This project was further supported by the European Research Council (ERC StG Strong-Q, Grant 676842), the Foundation for Fundamental Research on Matter (FOM) Projectruimte grants (15PR3210, 16PR1054) and by The Netherlands Organisation for Scientific Research (NWO/OCW), as part of the Frontiers of Nanoscience program and through a Vidi Grant (Project No.\ 680-47-541/994).

\setcounter{figure}{0}
\renewcommand{\thefigure}{S\arabic{figure}}
\setcounter{equation}{0}
\renewcommand{\theequation}{S\arabic{equation}}

\clearpage

\section{Supplementary Information}

\subsection{Detailed fabrication, challenges and solutions}

\subsubsection{Detailed fabrication}
Our devices are fabricated in \SI{200}{nm} thick stoichiometric SiN deposited via low-pressure chemical-vapor deposition (LPCVD) on a plain \SI{200}{\mu m}-thick Si substrate. We then lithographically define a \SI{500}{nm} thick electron-beam sensitive resist (AR-P 6200.13) in the shape of our photonic crystal trampolines and transfer the pattern into the SiN device layer with a \ch{CHF3}/\ch{O2} plasma etch. In the case of our single photonic crystal trampolines the backside of the wafer is then patterned with square openings to fully etch through the entire Si wafer without forming a double membrane array. In the case of fabricating the latter, we first thoroughly strip the remaining electron beam resist with a suitable remover at elevated temperatures (Baker PRS-3000 at \SI{80}{\celsius}) to ensure a clean surface after the first pattern transfer step.
We then repeat the same procedure of transferring the trampoline pattern into the second device layer while protecting the already patterned front side as to minimize exposure of both device layers to the clean room environment. This cannot fully be avoided as both device layers will get in contact during the spin coating procedure, i.e.\ with both the spin coater chuck and the hot plate surface during tempering. Despite that fact, we do not see clear negative effects on neither the mechanical nor optical properties of our resonators.
After the pattern transfer into both SiN layers, we again clean the chip surfaces thoroughly from any organic compounds. We first use Baker PRS-3000 at \SI{80}{\celsius} to remove the remaining electron resist off the surface followed by a hot Piranha solution at \SI{110}{\celsius}.
To release the trampolines, the chips are briefly rinsed in various water baths and then transferred to a \SI{30}{\percent} potassium hydroxide (KOH) solution at \SI{75}{\celsius}. The silicon is etched through the entire wafer for about two hours at a rate of \SI{1}{\mu m/min}. After the release, a 10 min hydrochloric acid (HCl) etch cleans off KOH residues of the exposed resonators surfaces. We then carefully transfer them into subsequent rinsing baths of water and isopropyl alcohol (IPA) before drying them in a critical point dryer (CPD) to avoid their exposure to viscous forces and surface tension.

\subsubsection{Fabrication yield}
We have found that by patterning the entire central pad with a PhC, even on its edges, as shown in the zoom-in of Figure~\ref{fig:device}c, the fabrication yield increases considerably. Devices with round PhCs as close as \SI{5}{\mu m} to the pad edge show either cracks or even fully break. Increasing the PhC diameter such that we cover more of the central pad with etch holes seems to reduce part of the large stress on the membranes which presumably is causing their rupture during release. This allows us to explore a much wider range of possible design parameters with even larger pad sizes, significantly improving on challenges like alignment between both membranes related to finite aperture losses (see Section~\ref{sec:aperture}), or using bigger beam waists in an optical cavity.

\subsubsection{Alignment between front and back membrane}
We align front and backside using an optical microscope to determine the coordinates of the patterns to be written with respect to one corner of our chips. By using this method, we introduce uncertainties to the correct coordinates between front and backside, leading to misalignments between \SI{10}{\mu m} and \SI{30}{\mu m} with good reproducibility, effectively reducing the overlap between both mirrors. This could be significantly improved by using topological alignment markers reaching through the entire chip, e.g.\ by deep reactive ion etching (DRIE). This would lead to better alignment between both membranes with the drawback of adding additional fabrication steps.

\subsubsection{Operation in the high finesse transmissive regime}
\begin{figure}[t]
  \includegraphics{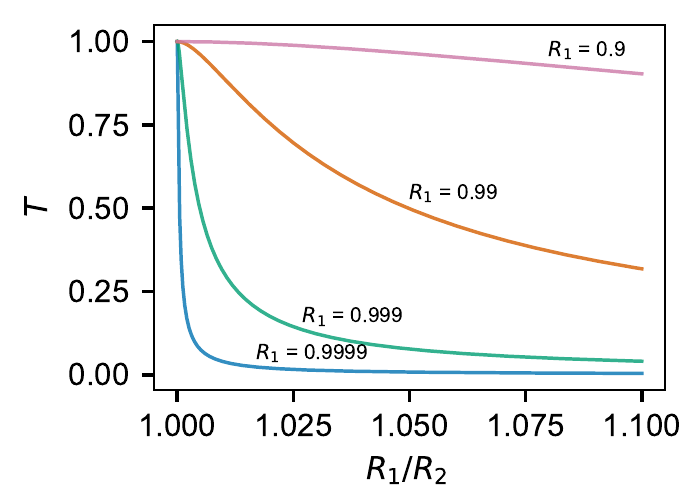}
  \caption{Influence of reflectivity mismatches between both mirrors. Plotted is the transmission $T$ versus the ratio of both mirror reflectivities $R_1/R_2$. For low finesse Fabry-P\'{e}rot cavities, bigger mismatches do not have as big of an influence on the the transmission as for the high finesse cases.}
  \label{fig:Impedance}
\end{figure}

In order to understand the importance of matching the reflectivities between both mirrors, we plot the theoretical transmission $T$ of the Fabry-P\'{e}rot cavity with respect to the ratio $R_1/R_2$ of its individual mirror reflectivities (see Figure~\ref{fig:Impedance}). The transmission (without losses) follows the following equation for normal incident light

\begin{equation*}
T = \frac{(1-R_1)(1-R_2)}{(1-\sqrt[]{R_1 R_2})^2}.
\end{equation*}

One can see that the transmission only reaches unity for matching mirror reflectivities and drops quicker the higher the finesse of the cavity becomes, i.e.\ for increasing $R_1$ and $R_2$. In the case of $R_1=\SI{90}{\percent}$, mismatches of up to \SI{10}{\percent} do not cause a big drop in transmission yet (pink curve, see also low and mid R transmission plots in Figure~\ref{fig:transmission}). For very high mirror reflectivities above $R_1=\SI{99.99}{\percent}$ even small mismatches between both mirrors already lead to a significant and rapid reduction in transmission (blue curve, Figure~\ref{fig:transmission}).

\begin{figure}[t]
  \includegraphics{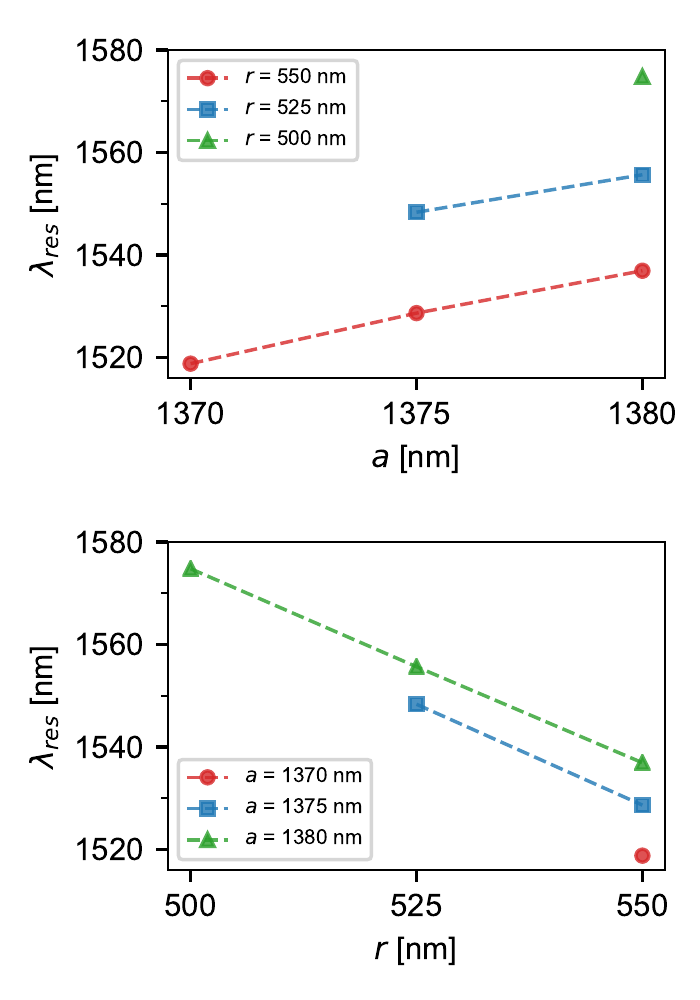}
  \caption{Influence of the PhC design parameters on its resonance wavelength $\lambda_\text{res}$. Plotted are measured resonance maxima for varying lattice constants $a$ and radii $r$. For increasing lattice constants, the maximum of the PhC resonance shifts to higher wavelengths, whereas it decreases for increasing radii.}
  \label{fig:Tuning}
\end{figure}

It is crucial to have good control over the tuning of our PhC resonances in order to account for mismatches between both mirrors induced by fabrication imperfections, especially in high finesse cavities. We therefore fabricated single membranes with various PhC parameters in order to see its influence on the maximum of their resonance $\lambda_\text{res}$ (see Figure~\ref{fig:Tuning}). We vary their radius $r$ as well as lattice constant $a$ and find a linear behavior around our operating wavelength of \SI{1550}{nm}. Taking the measured values of the lines with three data points, we can determine the slopes to be $\Delta \lambda_\text{res} \propto 1.81 \cdot \Delta a$ for a fixed radius of $r=\SI{550}{nm}$, and $\Delta \lambda_\text{res} \propto -0.76 \cdot \Delta r$ for a fixed lattice constant of $a=\SI{1380}{nm}$.

In order to test how much the reduced dip depths can be attributed to mismatching mirror reflectivities, we fabricated double membrane arrays with varying PhC design parameters. We keep the devices on one side of the chip fixed while sweeping the resonances on the other side by $\pm$\ \SI{1.5}{nm} in their lattice constants $a$, effectively tuning $\lambda_\text{res}$ by more than \SI{5}{nm}. We found that we could increase the dip depth of the high R array resonances from \SI{1}{\percent} to up to \SI{10}{\percent}. Further, more finely spaced sweeps of these parameters should allow for even larger dip depths, while operating in a regime of high finesse.

\subsection{Setups}

\subsubsection{Optical characterization}
We obtain the optical spectrum of the devices using a laser tunable from \num{1500} to \SI{1630}{nm}. This laser beam is split into a path that is incident on the sample and a reference path, directed onto a photodetector. This serves to correct for drifts in the setup before the interaction with the sample. The incident beam is focused onto the sample, resulting in a beam waist of \SI{50}{\mu m}. Light transmitted through the sample is measured on a third photodetector. Light reflected from the sample follows the same optical path as the incident beam. We split these two, using a combination of a polarization beam splitter and a quarter-wave plate, and send the reflected beam to a separate photodetector. Before measuring a PhC, we calibrate the setup using a commercial dielectric mirror with a known reflectivity. After, the PhC is placed in the setup, its tip, tilt and position in relation to the beam waist is carefully adjusted, and its measurements are compared to those of the dielectric mirror in order to obtain the transmission and reflection coefficients as a function of wavelength.

\subsubsection{Mechanical characterization}
We characterize the resonators' mechanical properties using a fiber-based homodyne interferometer. A laser beam is split into signal and local oscillator (LO) paths. The signal is focused onto the center pad of the trampoline resonators. Light reflected from the devices is split from the incident beam path using a fiber circulator and then combined with the LO using a tunable fiber coupler, in order to precisely set the coupling ratio to 50:50. The coupler output is measured using a balanced detector. The low frequency output of the detector is used to lock the phase of the signal and LO, using a PID controller and a fiber stretcher which is connected to the LO path. Locking the phase between the LO and back-reflected signal beam on the phase quadrature allows us to be sensitive to the displacement of our mechanical resonators. The high frequency detector output contains the information we are interested in retrieving, and is fed into a Spectrum Analyzer (SA). Finally, a piezo actuator attached to the stage can drive the mechanical modes of interest, whose mechanical quality factors are then determined by means of ring-down measurements.

\subsubsection{Optomechanical characterization}
To characterize the optomechanical properties of the devices we place them in the center of a rigid optical cavity, composed of two commercial mirrors which are \SI{48.1}{mm} apart. The mirrors are curved with an equal radius of \SI{25}{mm}, making a stable cavity with a ${\rm FSR_c} = \SI{3.12}{GHz}$, an empty cavity half-width at half-maximum linewidth of \SI{550}{kHz} and a corresponding finesse of \num{3000}. We estimate the cavity waist to be \SI{49}{\mu m}, considerably smaller than the PhC diameter of \SI{300}{\mu m} and approximately the same as the one used on the optical characterization setup.

When empty, we align the incident laser to the cavity, achieving a mode matching to the $\rm TEM_{00}$ modes larger than \SI{90}{\percent}. By measuring light that is both transmitted and reflected from the cavity and by scanning the laser frequency, previously calibrated using a wavemeter, we obtain the cavity spectrum.

We then position the membrane inside the cavity using a 3-axis piezoelectric stick-slip positioner. This allows us not only to precisely align the PhC membranes to the cavity waist, but also to probe multiple devices on the same chip. The positioner is mounted on top of a stage which enables the alignment of the membranes' tip and tilt in relation to the cavity axis. The x-axis positioner can also be operated in a conventional continuous voltage mode, which lets us displace the membrane by up to \SI{5}{\micro m} along the cavity axis.

Finally, for a given laser wavelength and position we scan the laser frequency and measure the transmitted power. The maxima of the transmission correspond to cavity resonances. We obtain the dispersion maps of Fig.~\ref{fig:dispersion} by repeating this measurement for multiple positions and wavelengths. Due to their low signal to noise ratio, in order to make the plots of Fig.~\ref{fig:dispersion} clearer, we apply a bandpass-pass filter to the data, ensuring it has no influence to the height and width of the resonances.

We would like to point out that on the optical characterization setup, due to the procedure we follow to do coarse wavelength sweeps, the wavelength has an uncertainty of \SI{0.5}{nm}, while the wavelength on the optomechanical characterization setup is much better defined. To compensate for this mismatch, we use the coupling minima as a reference for the wavelength where the $\sqrt{R}$ minima should occur in Fig.~\ref{fig:coupling} and shift $\sqrt{R}$ accordingly. To be specific, on that figure, the spectra of the Low, Mid and High R devices were shifted by \num{0.1}, \num{0.1} and \SI{-0.3}{nm}, respectively.

\subsection{Estimation of optical losses}\label{OpticalLosses}

\subsubsection{Single-membrane minimum transmission}
\label{sec:SM_transmission}

\begin{figure}[t]
	\centering
  \includegraphics{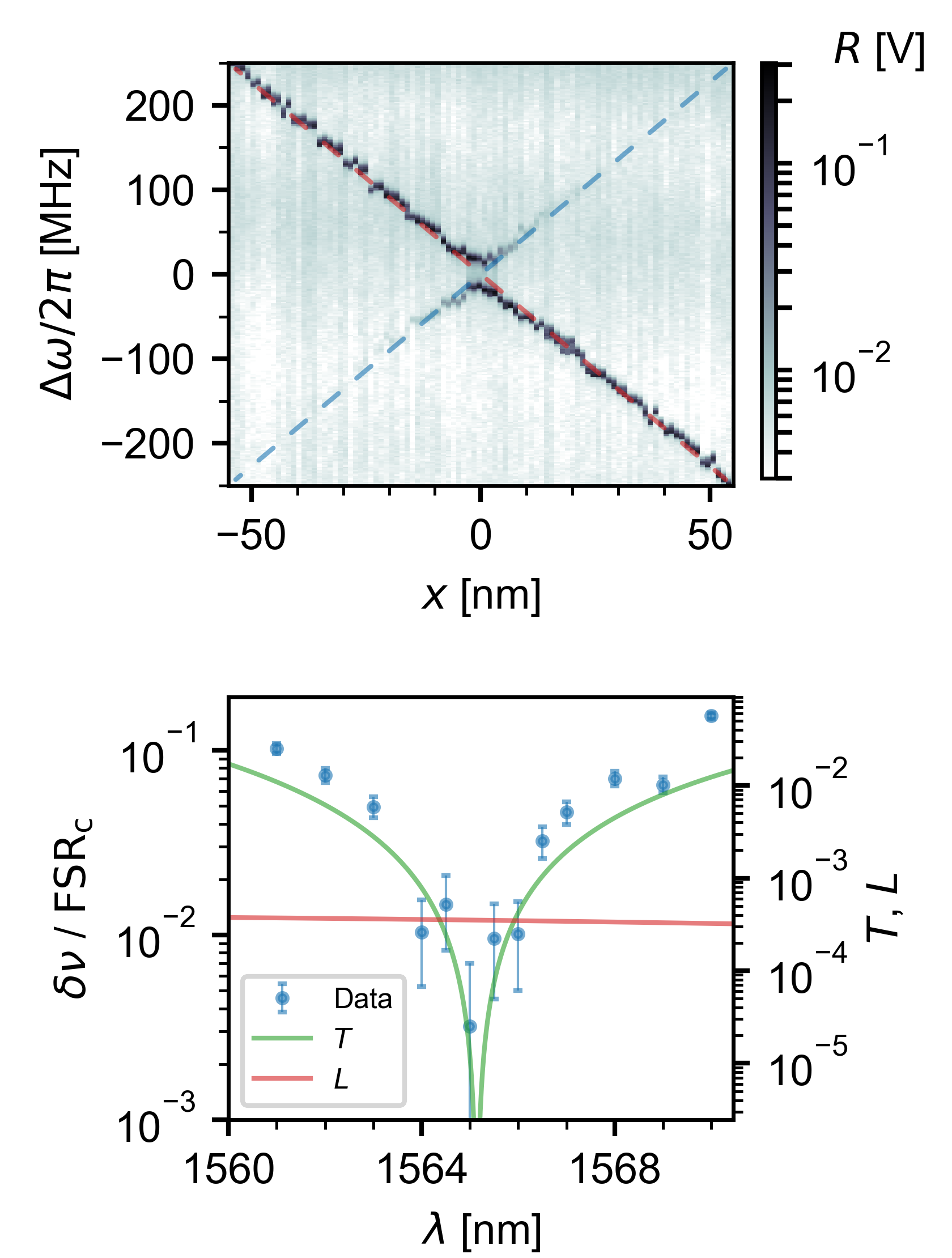}
  \caption{Top:\ Reflection of the optical cavity as a function of laser frequency $\Delta\omega$ and membrane position $x$ at a wavelength of \SI{1566}{nm}. Here, we study a sample with a PhC resonance at \SI{1565}{nm} and measure the splitting of the avoided crossing $\delta\nu$. Bottom:\ The points indicate the measured $\delta \nu$, normalized by the cavity free spectral range $\rm FSR_c$, as a function of the laser wavelength, close to the PhC resonance. These data can be converted into a membrane transmission, which is indicated on the right axis. The traces are the result of an $S^4$ simulation of the transmission (green) and absorption losses (red) of a single PhC membrane with similar geometry as the measured sample and an imaginary part of the refractive index of \num{1.9e-5}.}
  \label{fig:SM_highR_dispersion}
\end{figure}

To obtain a more accurate estimate of the maximum reflectivity achievable with our PhC membranes, we place a sample with a PhC resonance at \SI{1565}{nm} in the cavity setup described before. Fig.~\ref{fig:SM_highR_dispersion} (top) shows the cavity reflection as a function of laser frequency and membrane displacement $x$. The membrane divides the cavity into two half-cavities whose mode frequencies are a function of membrane displacement $x$. As $x$ increases, the length of the half-cavity above (below) the membrane increases (decreases), changing the mode frequency as indicated by the dashed red (blue) line. If the membrane was perfectly reflective, both half-cavity mode frequencies would become degenerate at a particular $x$. Realistically, the membrane has a non-zero transmission which allows some light to leak between the two half-cavities. This lifts the degeneracy and gives rise to an avoided crossing whose frequency splitting $\delta \nu$ can be used to extract the membrane transmission \cite{Stambaugh2014,Chen2017}.

We repeat this measurement for multiple wavelengths close to the PhC resonance. The frequency splittings normalized by the free spectral range are plotted as blue circles in Fig.~\ref{fig:SM_highR_dispersion} (bottom). The smallest $\delta \nu$/$\rm FSR_c$ measured for this sample was \num{3.2e-3} at \SI{1565}{nm}. Using the method of Stambaugh et al.~\cite{Stambaugh2014}, this corresponds to a a minimum PhC transmission of \num{2.5e-5}.

\subsubsection{Double membrane transmission loss}
In each round-trip, some light inside the double membrane etalon is transmitted through the PhC mirrors. If both membranes had exactly the same PhC resonance wavelength and if the highest finesse peak was exactly at the same wavelength as that resonance, this would result in a round-trip transmission of \num{5e-5}. However, the double-membrane peak is not, in general, at the PhC resonance. For a double-membrane with similar PhC resonances, the peak can be, at most, \SI{3}{nm} (approximately $\rm FSR_{DM}/2$) away from the PhC resonance. According to Fig.~\ref{fig:SM_highR_dispersion}, this sets an upper boundary to the round-trip transmission of \num{2.6e-2}.

\subsubsection{Material losses}
When light interacts with the SiN layer, some of it will be absorbed by the material or scattered away due to fabrication imperfections. To estimate the magnitude of these effects, we use $S^4$ to simulate the reflection and transmission through a PhC with similar parameters to those of Figure~\ref{fig:SM_highR_dispersion}. We have considered the material to have an imaginary part of the refractive index of \num{1.9e-5}~\cite{Stambaugh2014}, which accounts for not only absorption but also other loss mechanisms such as scattering~\cite{Chen2017}, and we calculate the losses as $L=1-R-T$, where $R$ is the reflection and $T$ the transmission coefficients. The simulation results are shown on the lower part of Fig.~\ref{fig:SM_highR_dispersion}. We see that the measured transmission follows the simulation quite well. Within this wavelength range, the losses are approximately constant and have a value of \num{3.5e-4}.

\subsubsection{Finite aperture loss} \label{sec:aperture}
Any Fabry-P\'{e}rot inteferometer with a finite aperture will lose some of the light through diffraction at the mirror edges~\cite{Siegman1986,Svelto2010}. These losses are higher for smaller mirrors and for increasing cavity stability parameter. In particular, a plane-parallel Fabry-P\'{e}rot cavity has a stability parameter $g=1$, which makes it particularly susceptible to finite aperture losses.

To estimate these, one can calculate the cavity Fresnel number $N=a^2/L\lambda$, where $a$ is the mirror radius and $L$ is the cavity length, and obtain the estimated losses per cavity round-trip from tables in literature~\cite{Siegman1986,Svelto2010}. Given a mirror diameter of \SI{260}{\mu m}, the Fresnel number of our devices is \num{54}, which corresponds to a loss per round-trip of \num{2e-3}.

Notice that this effect could be directly mitigated either by making the PhC membranes larger, or by controlling the wavefront of the field with one of the PhC, effectively realizing a focusing mirror~\cite{Guo2017}. This would reduce the stability parameter of the cavity, making it less susceptible to finite aperture losses.

\begin{figure}[t]
	\centering
	\includegraphics{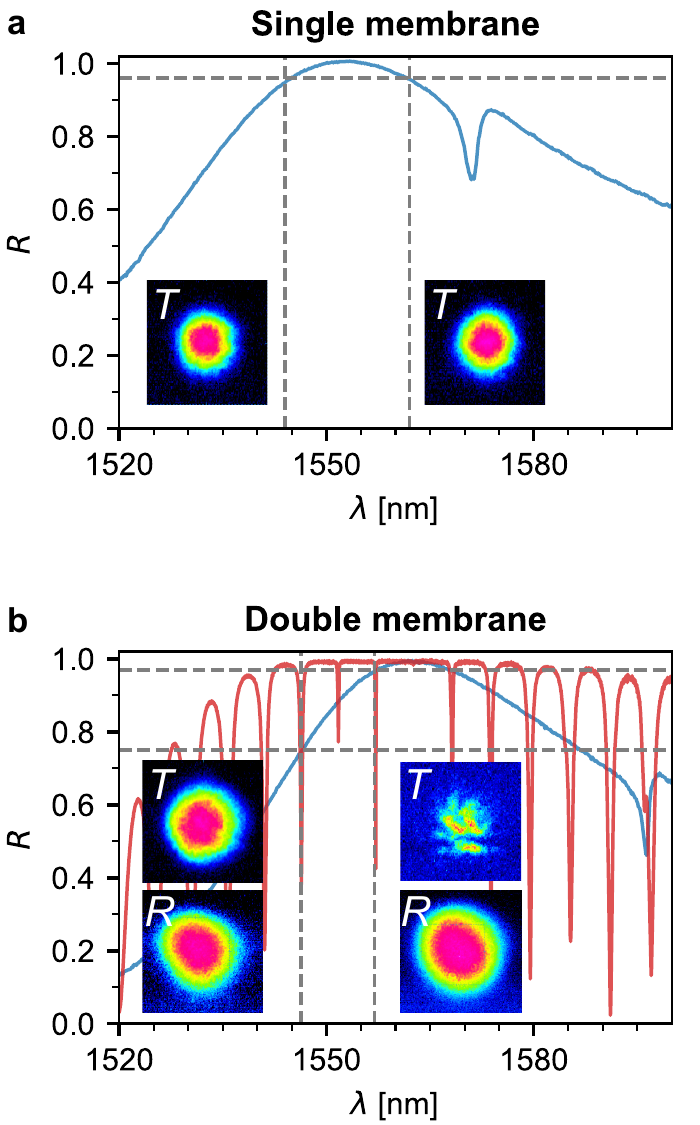}
	\caption{Shown are the reflection spectra, together with the transmitted ($T$) and reflected ($R$) beam mode profiles of single \textbf{a} and double membranes \textbf{b}. The dashed lines indicate the reflectivity and wavelength at which the mode profiles were measured.}
	\label{fig:mode_profiles}
\end{figure}

\subsubsection{Total losses}
Taking into account the previous results, we can estimate the total losses of the double-membrane array if its highest finesse peak is at the resonance wavelength of the PhC or if it is \SI{3}{nm} away from it. Using the finite aperture loss, the measured transmissions and the simulated material losses, we reach round-trip total losses of \num{2.8e-3} and \num{2.9e-2}, corresponding to cavity finesse values of \num{2243} and \num{217}. The lower bound of this range is in good agreement with the maximum finesse we measure in our devices of about \num{220}. However, most of our samples show maximum finesses which are approximately a factor of 5 smaller. This could be due to underestimations of scattering and diffraction losses, or due to additional absorption by material residues on the SiN layers or to the mismatch between the PhC resonances of both membranes.

\subsection{Mode profile analysis}
The reflection and transmission beam profiles can also help in understanding the behavior of our devices. We install flip mirrors in our setup which can send the optical beams to an IR-sensitive camera and record the beam profiles for single and double membranes, shown in Figure~\ref{fig:mode_profiles}a and b, respectively.

For the single membranes we obtain the beam profiles slightly detuned from the maximal reflectivity, around \SI{95}{\percent}, as otherwise the transmission is below the sensitivity of the camera. The measured modes have an overlap of approx.\ \SI{83}{\percent} with a Gaussian distribution, highlighting that the PhC structures distort the transmitted optical beams only slightly.

In Figure~\ref{fig:mode_profiles}b we plot the reflection spectrum of a double membrane (red) which individual membranes have a spectrum similar to the one shown in blue. The resonance with the highest finesse occurs at \SI{1562}{nm}, however its low dip depth makes the mode difficult to measure with our camera. The adjacent resonance at \SI{1557}{nm} shows the second highest finesse ($F=144$), corresponding to single membrane reflectivities of around \SI{97}{\percent}. Here we are able to measure the beam profiles for the transmitted and reflected light. While the reflection is mostly unaffected, the transmitted beam appears distorted. As the single membrane transmission does not show such behavior, we suspect the distortion results partly from scattering losses, as described in section~\ref{sec:aperture}. This loss mechanism becomes more dominant as the number of cavity round-trips, i.e.\ the finesse, increases. Indeed, for the resonance at \SI{1546}{nm} with a lower finesse of only \num{21}, corresponding to a single element reflectivity of \SI{75}{\percent}, the transmitted and reflected beam profiles have an overlap with a Gaussian distribution of more than \SI{84}{\percent}.

In addition, we would also like to note that the tip/tilt alignment becomes more important in double membrane arrays with high finesse, since the incident beam has to be properly mode matched to the cavity. This is further complicated by the plane-parallel geometry of our PhC cavities and could therefore be another main contribution to the observed mode distortion. This problem could be ameliorated by making one of the PhC mirrors a so-called \emph{focusing} PhC~\cite{Guo2017}, which can decrease the cavity stability parameter, making the mode matching and alignment easier.

\end{document}